\documentclass[letterpaper]{article} % DO NOT CHANGE THIS
\usepackage{aaai2026}  % DO NOT CHANGE THIS
\usepackage{times}  % DO NOT CHANGE THIS
\usepackage{helvet}  % DO NOT CHANGE THIS
\usepackage{courier}  % DO NOT CHANGE THIS
\usepackage[hyphens]{url}  % DO NOT CHANGE THIS
\usepackage{graphicx} % DO NOT CHANGE THIS
\usepackage{natbib}  % DO NOT CHANGE THIS AND DO NOT ADD ANY OPTIONS TO IT
\usepackage{caption} % DO NOT CHANGE THIS AND DO NOT ADD ANY OPTIONS TO IT
\usepackage{algorithm}
\usepackage{algorithmic}
\usepackage{booktabs}
\usepackage{newfloat}
\usepackage{listings}
\usepackage{algorithm}
\usepackage{algorithmic}
\usepackage{xcolor}
\DeclareCaptionStyle{ruled}{labelfont=normalfont,labelsep=colon,strut=off} % DO NOT CHANGE THIS
\floatstyle{ruled}
\newfloat{listing}{tb}{lst}{}
\floatname{listing}{Listing}
\title{AdaptJobRec: Enhancing Conversational Career Recommendation through an LLM‑Powered Agentic System}

\author{
    Qixin Wang\textsuperscript{\rm 1,2},
    Dawei Wang\textsuperscript{\rm 1},
    Kun Chen\textsuperscript{\rm 1},
    Yaowei Hu\textsuperscript{\rm 1}\\
    Puneet Girdhar\textsuperscript{\rm 1},
    Ruoteng Wang\textsuperscript{\rm 1},
    Aadesh Gupta\textsuperscript{\rm 1},
    Chaitanya Devella\textsuperscript{\rm 1}\\
    Wenlai Guo\textsuperscript{\rm 1},
    Shangwen Huang\textsuperscript{\rm 1},
    Bachir Aoun\textsuperscript{\rm 1},
    Greg Hayworth\textsuperscript{\rm 1}\\
    Han Li\textsuperscript{\rm 1}\thanks{corresponding author},
    Xintao Wu\textsuperscript{\rm 2}\footnotemark[1]
}
\affiliations{
    \textsuperscript{\rm 1}Walmart Global Tech\\
    han.li@walmart.com\\
    \textsuperscript{\rm 2}University of Arkansas\\
    xintaowu@uark.edu
}

\usepackage{bibentry}
\begin{document}

\maketitle

\begin{abstract}
In recent years, recommendation systems have evolved from providing a single list of recommendations to offering a comprehensive suite of topic-focused services. To better accomplish this task, conversational recommendation systems (CRS) have progressed from basic retrieval-augmented LLM generation to agentic systems with advanced reasoning and self-correction capabilities. However, agentic systems come with notable response latency—a longstanding challenge for conversational recommendation systems. To balance the trade-off between handling complex queries and minimizing latency, we propose AdaptJobRec, the first conversational job recommendation system that leverages autonomous agent to integrate personalized recommendation algorithm tools. The system employs a user query complexity identification  mechanism to minimize response latency. For straightforward queries, the agent directly selects the appropriate tool for rapid responses. For complex queries, the agent uses the memory processing module to filter chat history for relevant content, then passes the results to the intelligent task decomposition planner, and finally executes the tasks using personalized recommendation tools. Evaluation on Walmart’s real-world career recommendation scenarios demonstrates that AdaptJobRec reduces average response latency by up to 53.3\% compared to competitive baselines, while significantly improving recommendation accuracy.
\end{abstract}

%Evaluation on Walmart’s real-world career recommendation scenarios demonstrates that AdaptJobRec reduces average response latency by up to 53.3\% compared to competitive baselines, while significantly ($p < 0.01$) improving recommendation accuracy on both Hit@10 and NDCG@10.

% Uncomment the following to link to your code, datasets, an extended version or similar.
% You must keep this block between (not within) the abstract and the main body of the paper.
%\begin{links}
%    \link{Code}{https://aaai.org/example/code}
%    \link{Datasets}{https://aaai.org/example/datasets}
%    \link{Extended version}{https://aaai.org/example/extended-version}
%\end{links}

\section{Introduction}
In recent years, the complexity of conversational recommendation systems (CRS) has been continuously increasing~\cite{zhang2024towards, huang2023recommender, fang2024multi}, evolving from offering a single type of recommendation to providing a range of services on specific topics. To improve task completion rates, these systems have progressed from LLM-based retrieval-augmented generation (RAG) models~\cite{friedman2023leveraging, gao2023chat, liu2023conversational, kuzi2024bridging} to sophisticated agentic systems with advanced reasoning and self-correction capabilities. Such systems dynamically integrate multiple modules (e.g., planning and memory processing) and various tools (e.g., databases, search engines, and knowledge graphs) to better serve users’ needs~\cite{zhang2024towards}. For instance, the ReAct architecture~\cite{yao2023react}, developed by Princeton University and the Google Brain team, is a widely used agentic architecture in the industry. It employs a reasoning agent that leverages Chain-of-Thought (CoT) to decompose a user query into multiple sub-tasks, invoking corresponding tools to obtain information and complete tasks based on the feedback step by step. Another widely used architecture, the Plan and Execute agent~\cite{wang2023plan} incorporates a separate replan module to assess task completion quality and generate the subsequent steps, enhancing the model’s reliability in handling complex problems. Other innovative approaches include the RecMind agentic recommendation system~\cite{wang2023recmind} designed by the Amazon Alexa AI team, which proposes a novel Self-Inspiring Planning module that considers previously explored states in Tree of Thought (ToT)  while planning the next step, effectively improving the planning ability of the agentic system, and the MACRS Agentic system \cite{fang2024multi}, which employs an Asking Responder Agent to elicit user preferences through additional dialogue rounds before generating recommendations. Meanwhile, InteRecAgent~\cite{huang2023recommender} proposed by Microsoft Research Asia improves long-conversation handling with components such as a memory bus, dynamic demonstration-augmented task planning, and reflective processing.

\begin{figure*}[t!]
  \centering
  \includegraphics[width=1\linewidth]{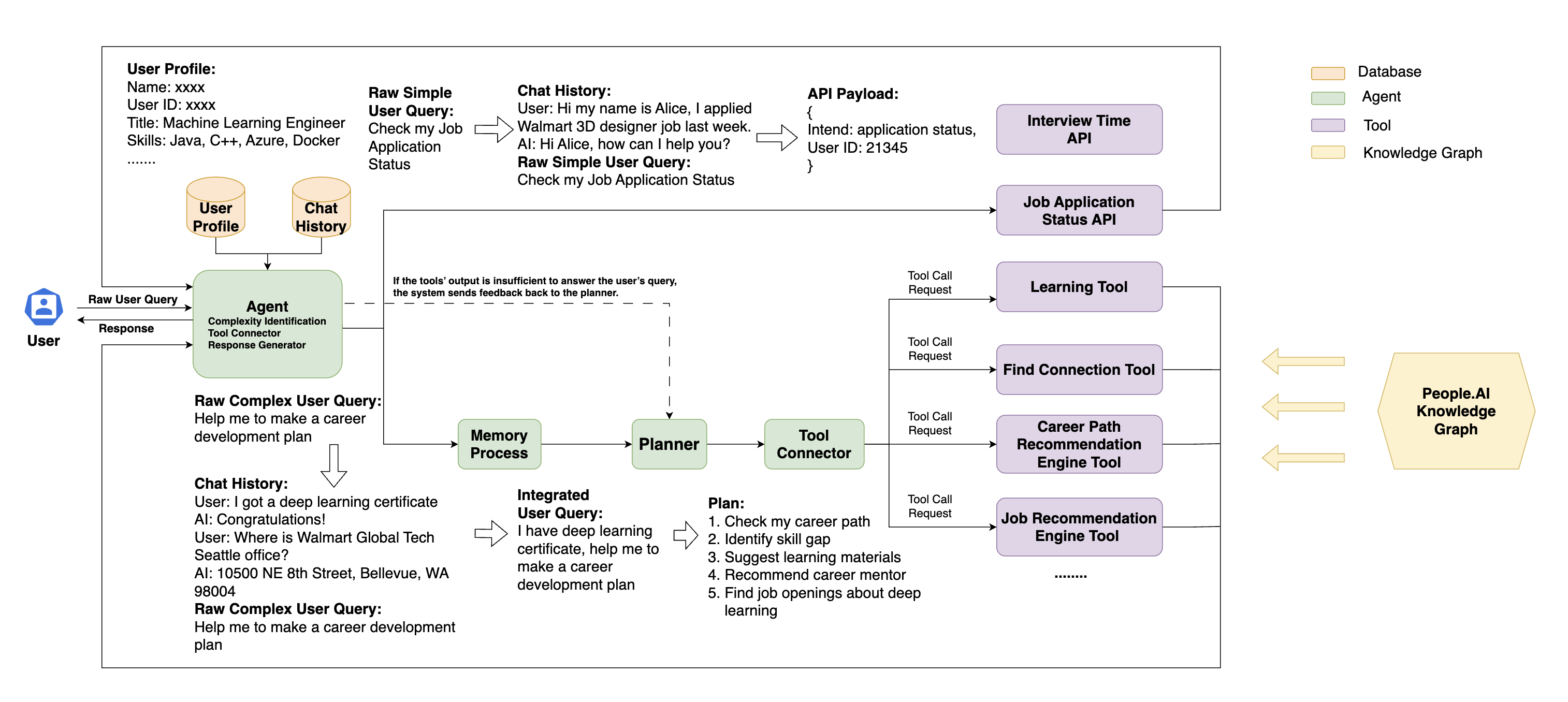}
  \caption{Architecture of AdaptJobRec Agentic System}
  %\Description{Architecture of AdaptJobRec Agentic System.}
  \label{fig:Arch_Adapt}
\end{figure*}

Although these planning, memory, and replan modules significantly enhance an agentic system's reasoning capability and problem-solving ability, they also increase response latency. To address this challenge, researchers in JD eCommence proposed the concept of first token latency \cite{nie2024hybrid}, demonstrating that an agentic CRS supported by a fine-tuned LLM can reduce the time required to output the initial response token, making an important contribution in this area. 

%To address the aforementioned issues, we propose AdaptJobRec, an agentic job recommendation system that integrates a complexity identification mechanism to distinguish between simple and complex user queries. For simple queries, it provides direct and quick responses, while for complex queries, it employs memory processing and task decomposition planning to deliver more precise answers. In addition, we have implemented three personalized tools—Job Recommendation Tool, Career Path Growth Tool, and CyperQL Tool—which are integrated into the system to enhance its capabilities. By comparing our proposed method with several baselines on the real Walmart associate conversation dataset, we found that our approach not only achieved better job recommendation performance but also required fewer conversation turns and less time.

%\footnote{We use Walmart as a placeholder for the real company name to maintain anonymity.}

Solving the response latency issue is particularly critical in contexts like Walmart's conversational job recommendation system. Walmart is a leading Fortune 500 company with millions of job applicants per year, where many users require rapid access to simple information, such as application status or interview time. In such cases, users are concerned with when they receive the needed information, merely reducing first token latency is insufficient. To overcome the delays of agentic systems, we propose the AdaptJobRec architecture. This architecture integrates a complexity identification mechanism that distinguishes between simple and complex user queries. It employs the memory processing and task decomposition planner exclusively for complex queries, ensuring rapid responses for simple queries while maintaining the capability to handle complex user needs. We compare AdaptJobRec against  two fine-tuned LLMs career path models~\cite{liu2024deepseek, touvron2023llama} and three widely used agentic systems~\cite{wang2023plan, yao2023react, fang2024multi}, demonstrating that AdaptJobRec outperforms these methods in prediction accuracy while delivering lower response latency.

Furthermore, besides reducing response latency in conversational recommendation systems, an equally important goal is minimizing the number of dialogue rounds needed for users to obtain key information. To achieve this goal, we develop a novel few-shot memory processing module that filters chat history for content relevant to the current user query, eliminating redundant planning, boosting planner accuracy, and enabling more precise tool selection. We also developed an intelligent task decomposition planner capable of generating a nested sub-task list, grouping tasks that can be executed asynchronously into the same sub-list. The AdaptJobRec agentic conversational recommendation system integrates memory processing, a task decomposition planner, and personalized recommendation tools based on user profiles and behavior. These components reduce the ask-back rounds needed to clarify user needs and the number of follow-up queries after receiving ineffective recommendations.

\section{Methodology}

\subsection{Architecture of AdaptJobRec}

As depicted in Figure~\ref{fig:Arch_Adapt}, AdaptJobRec comprises several key components: an agent with a complexity identification mechanism, a few-shot learning memory process module, a task decomposition planner, and recommendation tools powered by the Walmart People.AI knowledge graph. This knowledge graph includes 1.6 million nodes and 83 million edges, representing entities such as job titles, openings, associates, applicants, and skills etc. This knowledge graph serves as the foundation for a series of personalized recommendation tools. The AdaptJobRec agent quickly addresses simple queries using different tools, while complex queries trigger the memory process module and task decomposition planner to activate a suite of personalized recommendation tools to generate a high quality response.

In conversational recommendation systems, a user's intent is rarely fully captured from their initial query. To deliver more accurate recommendations, it is essential to consider not only the query itself but also the user profile and behavior. For instance, consider a senior software engineer with six years of Java programming experience who states, ``Find me some manager positions in the Seattle area; if none are available, other big cities will do." In this case, suggesting a Walmart store manager role in Seattle would be inappropriate, while recommending a lead software engineer position in New York that requires over five years of Java experience would better meet the user's needs, even if the former is closer to the original user query in the semantic vector space. Moreover, taking user behavior into account can further enhance recommendation quality. For example, if a senior software engineer recently saved three eCommerce related job openings on the Walmart careers website, it strongly indicates an interest in that field. Even without an explicit mention in the conversation, the recommendation system should proportionally increase the presence of eCommerce job openings in its results. Based on this idea, several tools in AdaptJobRec have been designed as personalized recommendation engines powered by the People.AI knowledge graph that replaces traditional search engines.

\subsection{Complexity identification mechanism}
\label{sec:Comp} 
The Walmart job conversational recommendation system categorizes user queries into two types. The first type includes simple and clear queries, such as ``help me check job application status", which require a fast and straightforward response. The second type involves more complex or ambiguous queries, such as ``can you create a career development plan for me?",  which should be decomposed into multiple sub-tasks: 1) Recommend a career path based on the user’s current position, skills, and career interests; 2) Identify the skill gap between the user and the potential next role; 3) Suggest learning resources based on the identified skill gap; 4) Recommend a career mentor based on the skill gap and potential next role; 5) Suggest job openings that the user can apply for. The system then invokes the necessary tools to complete each sub-task and synthesizes the results to generate a comprehensive response.

\begin{figure}[h]
  \centering
  \includegraphics[width=1\linewidth]{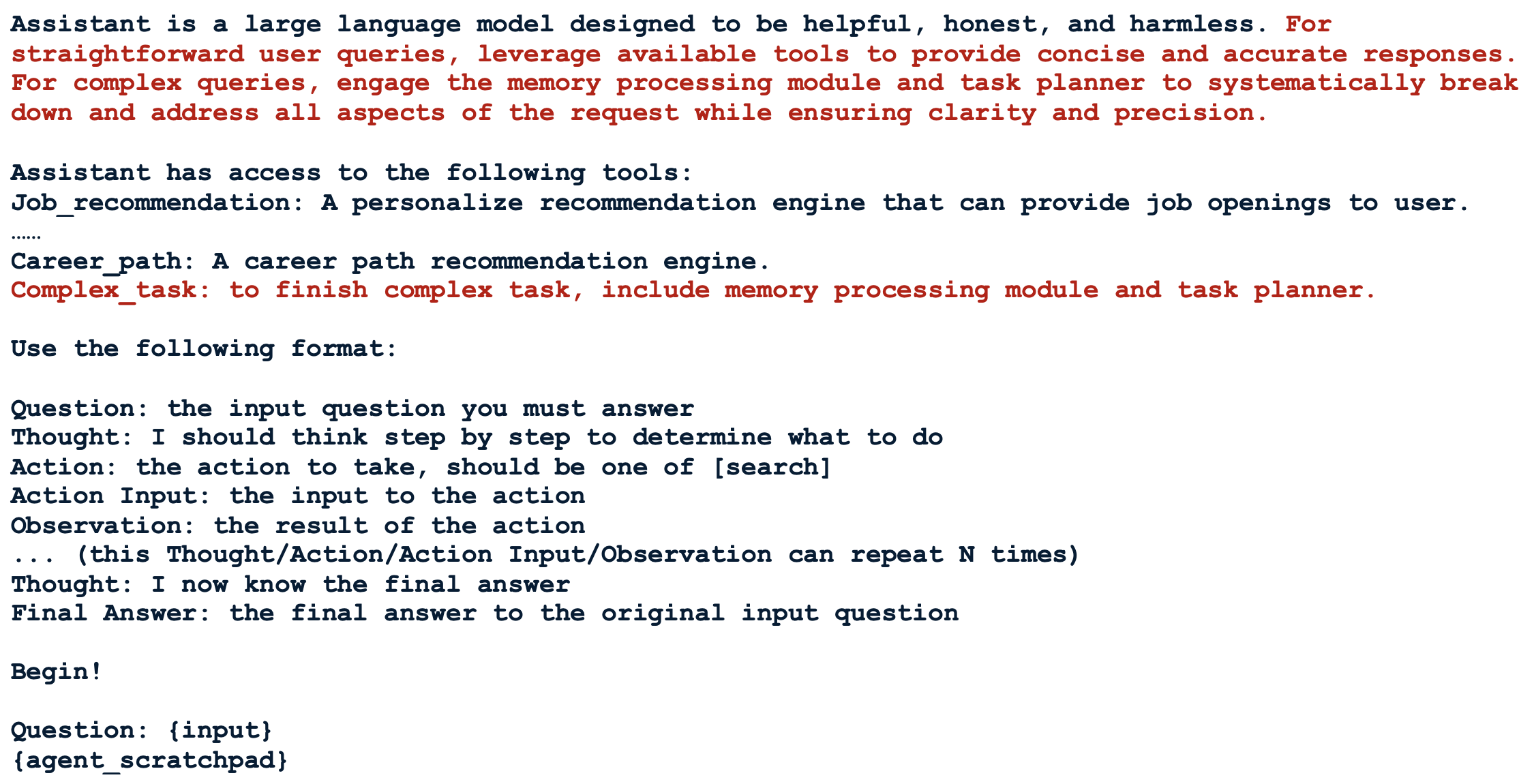}
  \caption{System Prompt of Complexity Identification Mechanism.}
  %\Description{System Prompt of  Complexity Identification Mechanism}
  \label{fig:sys_prompt}
\end{figure}

If we were to handle all types of user queries using an agent that includes modules such as a planner and a memory process module, the agentic CRS would respond too slowly to the simpler user queries, negatively impacting user experience. Therefore, we have designed an agentic system with a user query complexity identification mechanism, which is implemented using a customized system prompt. As shown in Figure~\ref{fig:sys_prompt}, the key parts of this system prompt that enable it to distinguish between simple and complex queries are highlighted in red. When a simple user query is received, the agent merges the user query with chat history without additional process and directly leverages tools such as the personalized recommendation APIs, or the People.AI knowledge graph to respond rapidly. For a complex query, the agent forwards the user query and additional relevant information (such as chat history, user profile, and career interests) to a few-shot learning memory processing module that integrates only the relevant memories to create an enhanced query, which is then decomposed into sub-tasks by a planner. The agentic system then employs the appropriate tools corresponding to tasks to fetch information, such as invoking the job recommendation engine, career path recommendation engine, or converting a sub-task into a Cypher query for the People.AI knowledge graph, and ultimately synthesizes them to provide a comprehensive response. When the information provided by the tool is not sufficient to satisfy the user’s query, the system sends feedback to the planner, initiates a new planning cycle, and repeats the tool selection process. This process is indicated by the dashed arrow in Figure~\ref{fig:Arch_Adapt}. This optimization can significantly reduce the latency of the LLM agentic system for simple queries while ensuring the accuracy of responses to complex and challenging user queries. 

%We also developed an Information Checker module to validate tool outputs. If the information is insufficient to response the user's query, it sends feedback to the planner and triggers another round of planning. These optimization can significantly reduce the latency of the LLM agentic system for simple inquiries while ensuring the accuracy of responses to complex and challenging user queries. 

%\begin{itemize}
%\item {\bfseries xx}: xxx
%\end{itemize}

\subsection{Memory processing module}
While techniques like windowing~\cite{dai2019transformer}, summarization~\cite{wang2025recursively}, and VectorDB~\cite{hatalis2023memory} effectively compress  chat history and provide the necessary context for a conversation, they often omit important details or retain irrelevant content. Such imprecision undermines the planner’s ability to decompose complex queries in an agentic system. To improve the handling of complex queries in the AdaptJobRec agentic system, we design a series of few-shot learning examples covering various scenarios to build a memory processing module  that enables LLM to precisely extract chat history segments relevant to the current user query. 

\begin{figure}[h]
  \centering
  \includegraphics[width=1\linewidth]{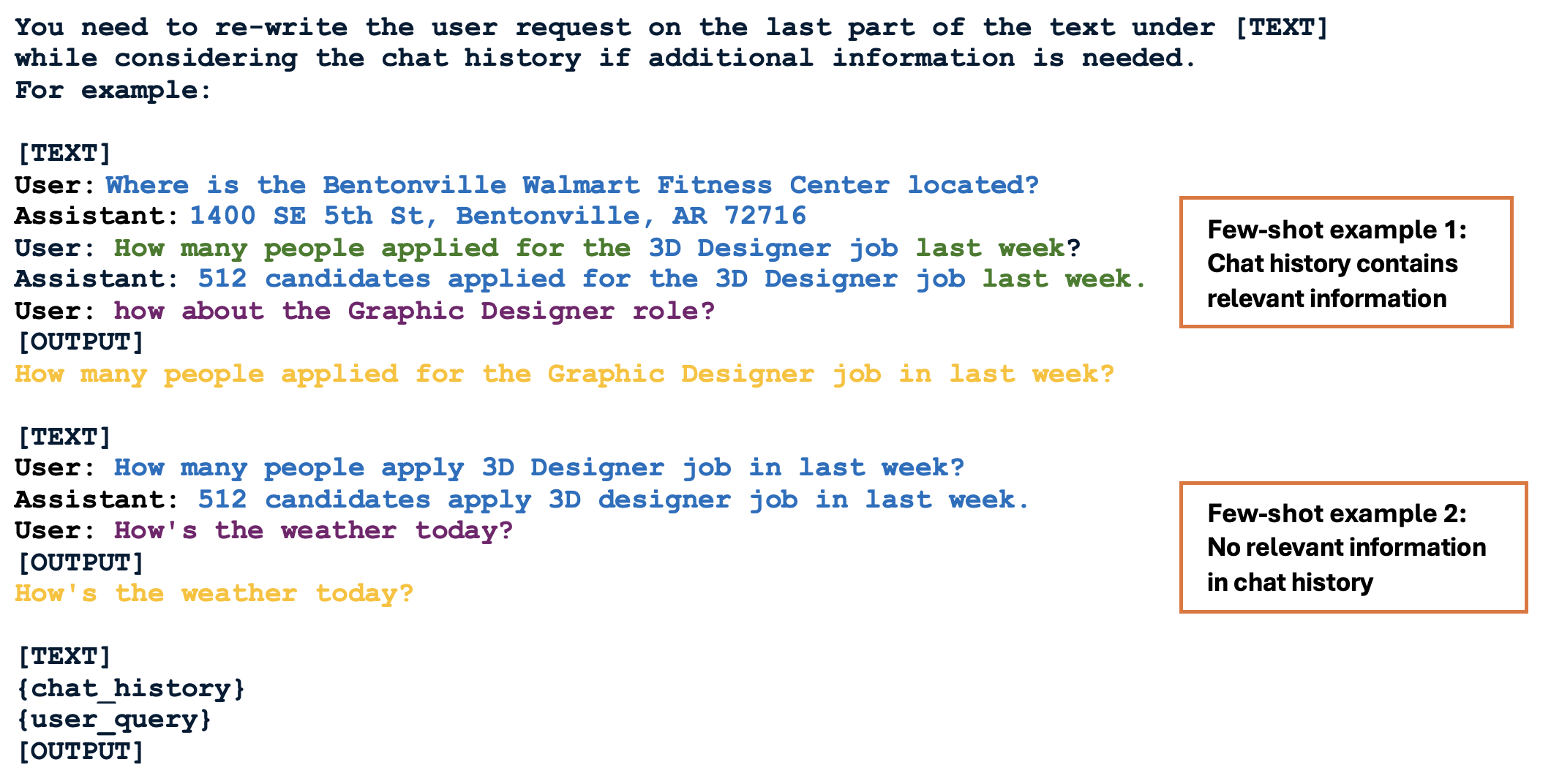}
  \caption{System Prompt of Few-shot Learning Memory Processing Module.}
  %\Description{System Prompt of few-shot learning memory processing module}
  \label{fig:mem_pro}
\end{figure}

Figure~\ref{fig:mem_pro} presents parts of the system prompt used by the memory process module. In few-shot example 1, chat history segments relevant to the current user query (highlighted in green) are extracted and merged with the current user query (highlighted in purple) under the [TEXT] section (irrelevant segments appear in blue). The resulting Integrated User Query is shown in orange after [OUTPUT]. Few-shot example 2 illustrates a scenario where the chat history contains no content pertinent to the current query.

\subsection{Planner}
We leverage the planner, which is a commonly used component in LLM-powered agentic systems, to decompose Integrated User Query (from the memory processing module) into sub-tasks and execute them with appropriate tools. In existing LLM-powered agentic systems ~\cite{wang2023plan, wang2023recmind} such as the Plan \& Execute Agent and RecMind Agent, planners typically generate a strictly ordered sequence of sub-tasks, invoking tools sequentially. However, while certain tasks inherently require sequential execution, many can be executed asynchronously, offering the potential to reduce response latency.

For example, given the query ``Which city has more machine learning engineer job openings, Seattle or Sunnyvale?”, the planner may generate the following task sequence:
\begin{itemize}
\item $[$`Get machine learning engineer job opening number from Seattle', `Get machine learning engineer job opening number from Sunnyvale'$]$
\item $[$`Compare job opening numbers of Sunnyvale and Seattle'$]$
\end{itemize}

The first two retrieval operations can be executed in parallel, followed by the comparison step. 

\begin{figure}[h]
  \centering
  \includegraphics[width=1\linewidth]{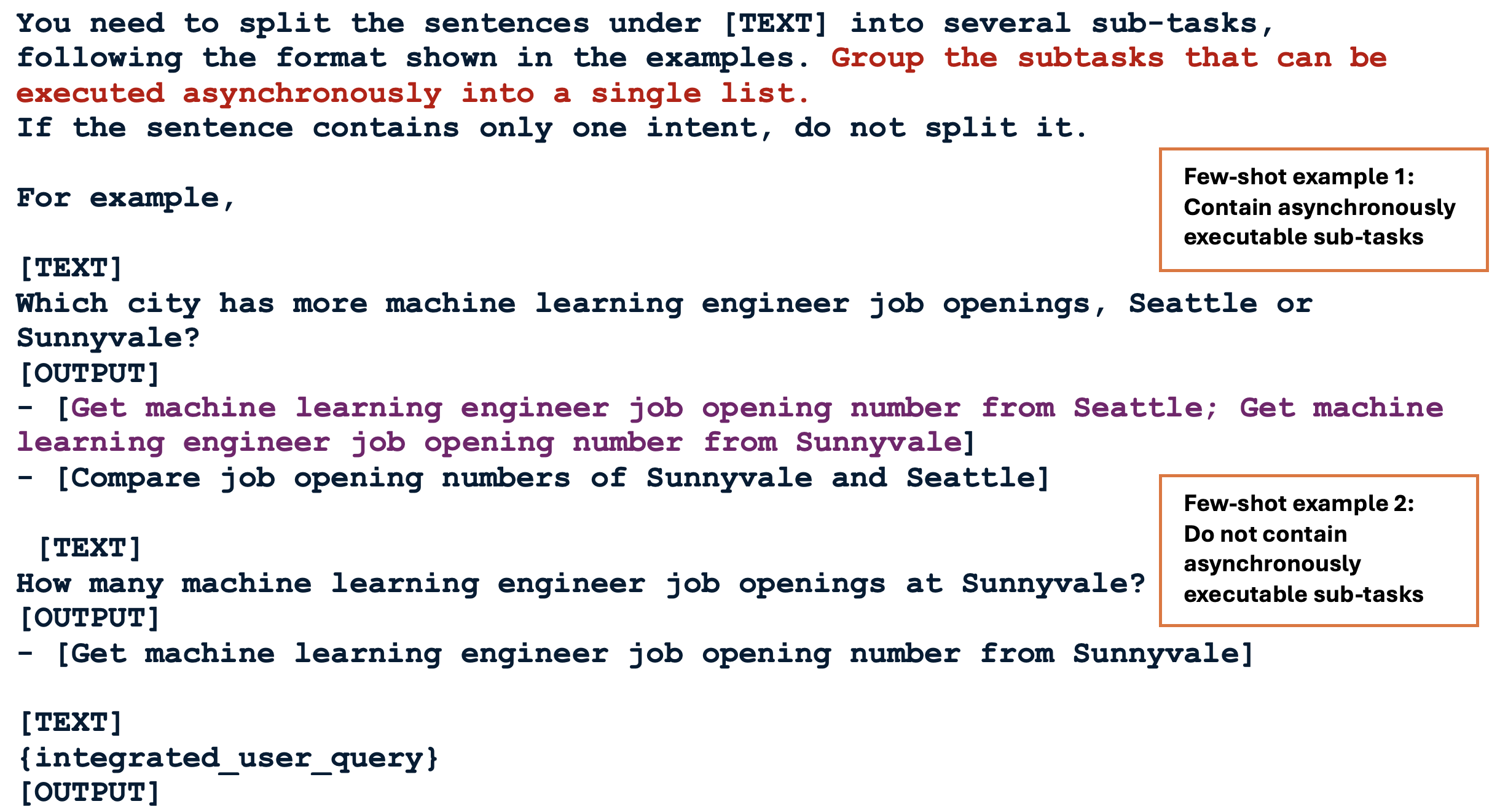}
  \caption{System Prompt of AdaptJobRec Planner.}
  %\Description{System Prompt of  Complexity Identification Mechanism}
  \label{fig:sys_prompt_planner}
\end{figure}

Therefore, in AdaptJobRec, we enhance the existing planner to output a nested list, grouping asynchronously executable sub-tasks within the same sub-list. This functionality is implemented via a few-shot learning module, with part of the system prompt shown in Figure~\ref{fig:sys_prompt_planner}. The prompt explicitly instructs the LLM to group sub-tasks that can be executed concurrently (highlighted in red). Few-shot example 1 shows an Integrated User Query containing two sub-tasks that can be executed in parallel (highlighted in purple). Few-shot example 2 illustrates a case with no asynchronously executable sub-tasks.

%(see \nameref{sec:Comp} for an example of such task decomposition).
%\xw{check the sec:Comp is undefined.}
%In AdaptJobRec agentic system, we implement LangGraph’s Planner module \cite{easin2024intelligent}, which combines fast response times with high quality planning. 

\subsection{Tools}

{\bfseries Personalized Job Recommendation Tool} When a user starts a conversation, the personalized job recommendation engine extracts their current job title, queries the People.AI knowledge graph for adjacent positions, and selects active openings with the most recent posting dates. It then matches key entities (e.g., skills, education, location) from the user profile to each candidate job opening, computes and sums similarity scores, and normalizes the sum score by the total number of entities. A user interest function then adjusts the normalized score based on real-time interactions (e.g., clicks, saves, likes, dislikes) with job openings across different job families. Both entity match weights and interest function parameters are tuned by Bayesian optimization on click data to maximize Click Through Rate (CTR), and the top 20 ranked results are forwarded to the main agent to generate the response.

%\label{sec:subsection}

{\bfseries Career Path Recommendation Tool} Career path recommendations fall into two categories. In one, the user explicitly states a desired destination, for example, ``I want to become a principal 3D designer; what should I do?" In the other, the destination is not specified, for instance, ``I just received a job offer for the Walmart Merchant position; please show me the future development prospects for this role." For the first scenario, our career path recommendation tool extracts the user’s current position from their profile and applies a shortest path algorithm using edge weights from the people.AI knowledge graph to identify and suggest potential career paths. For the second scenario, we develop a personalized career path growth algorithm that recommends several tailored career paths. This algorithm considers the user’s current position, skill set, and domain preferences, gleaned from recent activity, to comprehensively illustrate the role’s future development. 

%\subsection{CyperQL Tools} 

{\bfseries Cypher Tools} We develop two types of Cypher Tools. The first consists of predefined Cypher code templates, where the agent selects the appropriate tool based on the user's intent and fills in key entities, minimizing response time for high frequency queries. The second is a more flexible Text-to-Cypher approach, which feeds both the knowledge graph schema and user input into the LLM to generate Cypher queries, allowing it to address a broader range of user queries.

%The Cyper Tools powered by People.AI knowledge graph transform users' natural language into executable CyperQL queries. It retrieves information from the People.AI knowledge graph and passes it to the agent. To minimize response time, we avoid relying on the LLM to generate CyperQL queries autonomously. Instead, we use a set of predefined query templates, selecting one based on the user's intent and filling in the key entities. This approach not only reduces the latency of our agentic recommendation system but also improves response accuracy.

%user CyperQL template response time shorter than GraphQL API
%\subsubsection{Subsubsection}
%\label{sec:subsubsection}
%This is a subsubsection.

\section{Application Deployment}

\begin{figure*}[t!]
  \centering
  \includegraphics[width=1\linewidth]{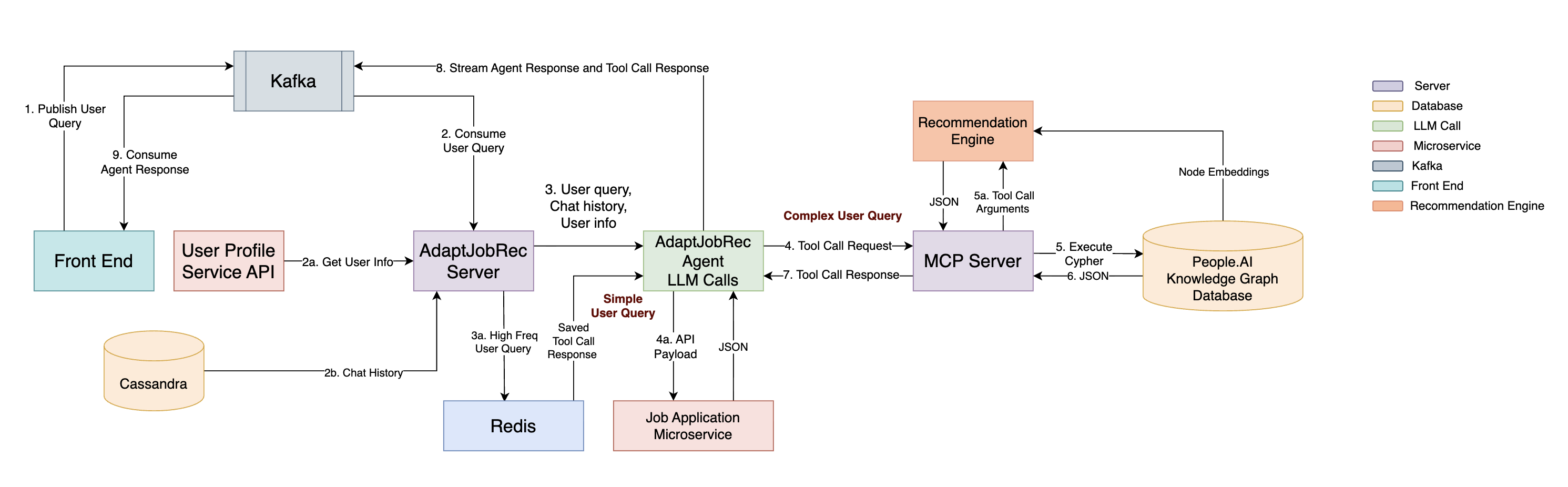}
  \caption{AdaptJobRec Agentic System Deployment Architecture}
  %\Description{Architecture of AdaptJobRec Agentic System.}
  \label{fig:Adapt_Deploy}
\end{figure*}

%As shown in Figure~\ref{fig:Adapt_Deploy}, AdaptJobRec is implemented as a set of independently deployable services. 
%\xw{It will be helpful to give an overview of components in Figure 4.}
%Cross service communication between the Front End and the AdaptJobRec agent component is facilitated through Kafka topics, while integration with the People.AI knowledge graph is established via the Model Context Protocol (MCP)~\cite{hou2025model}.

As shown in Figure \ref{fig:Adapt_Deploy}, AdaptJobRec is implemented as a set of independently deployable services, each responsible for a distinct function in query handling and recommendation generation. The Front End provides the user interface and communicates with the backend via Kafka, which serves as the streaming backbone for publishing user queries and delivering agent responses. The AdaptJobRec Server orchestrates backend operations, retrieving user information from the User Profile Service API and historical conversations from the Cassandra database, while leveraging Redis to cache frequently accessed query results and tool responses to reduce latency. The AdaptJobRec Agent hosts the LLM-based reasoning and decision logic, which classifies queries into simple or complex. Simple queries are routed to the Job Application Microservice, while complex queries are handled through the Model Context Protocol (MCP) ~\cite{hou2025model} server, which call recommendation engine tools or execute Cypher queries against the People.AI Knowledge Graph stored in the graph database.

Upon receiving a user query, the Front End publishes the corresponding topic to the Kafka cluster (1 in Figure \ref{fig:Adapt_Deploy}). The AdaptJobRec Server, subscribed to this topic, retrieves the query (2 in Figure \ref{fig:Adapt_Deploy}) and uses the login credentials provided by the Front End to call the User Profile Service API (2a in Figure \ref{fig:Adapt_Deploy}), obtaining personalized user information. In parallel, it fetches the user's conversation history from the Cassandra database (2b in Figure \ref{fig:Adapt_Deploy}).

To reduce response latency for high frequency queries (e.g., ``What can be the future role of a Walmart cashier?"), we integrate Redis with the AdaptJobRec Server to enable Cache Augmented Generation (3a in Figure \ref{fig:Adapt_Deploy}). This mechanism stores and reuses tool call responses for frequently asked queries, thereby eliminating redundant LLM calls and reducing overall response time.

Within the AdaptJobRec Agent, the complexity identification mechanism evaluates the user query, profile, and chat history (3 in Figure \ref{fig:Adapt_Deploy}) to classify the request as either simple or complex. For simple queries (4a in Figure \ref{fig:Adapt_Deploy}), the agent directly accesses the Job Application Microservice, which provides a suite of APIs to retrieve the information needed for response generation. For complex queries (4), the query and profile are routed through the memory process module and planner component before the agent extracts key entities and formulates a tool invocation request to the MCP Server. The MCP Server then either executes recommendation tools (5a in Figure \ref{fig:Adapt_Deploy}) or generates a Cypher query (5 in Figure \ref{fig:Adapt_Deploy}) to retrieve the required information from the People.AI Knowledge Graph (6 in Figure \ref{fig:Adapt_Deploy}), and returns the results to the agent (7 in Figure \ref{fig:Adapt_Deploy}).

Finally, the AdaptJobRec Agent synthesizes the final answer based on information retrieved from either the Job Application Microservice or the MCP Server and streams the response to Kafka (8 in Figure \ref{fig:Adapt_Deploy}). The Front End, acting as a Kafka consumer (9 in Figure \ref{fig:Adapt_Deploy}), processes this stream and delivers the final response to the user.

\section{Evaluation Result}

In this section, we evaluate the performance of AdaptJobRec on three tasks: Job Recommendation, Career Path Prediction, and a Pilot User Study using real-world data from Walmart company.

\subsection{Job Recommendation} 

\begin{table}[t]
  \centering
  \begin{tabular}{lccc}
    \toprule
    Methods & Hit@10 $\uparrow$ & NDCG@10 $\uparrow$ & MAP@10 $\uparrow$ \\
    \midrule
    RAG             & 0.2000 & 0.0391 & 0.0136 \\
    ReAct           & 0.2789 & 0.0747 & 0.0342 \\
    Plan \& Execute & 0.3127 & 0.0799 & 0.0364 \\
    MACRS           & 0.3120 & 0.0775 & 0.0350 \\
    \textbf{AdaptJobRec} & \textbf{0.3176} & \textbf{0.0810} & \textbf{0.0371} \\
    \bottomrule
  \end{tabular}
  \caption{Comparison on Job Recommendation Task}
  \label{tab:job}
\end{table}

\begin{table}[t]
  \centering
  \begin{tabular}{@{}lccc@{}}
    \toprule
    \shortstack[c]{AdaptJobRec\\ vs. Baseline} & \shortstack[c]{$p$-value\\(Hit@10)} & \shortstack[c]{$p$-value\\(NDCG@10)} &  \shortstack[c]{$p$-value\\(MAP@10)} \\
    \midrule
    RAG   & $<$0.001 & $<$0.001 & $<$0.001 \\
    ReAct  & $<$0.001 & $<$0.001 & $<$0.001 \\
    Plan \& Execute  & 0.001    & 0.008    & 0.027    \\
    MACRS & 0.001    & 0.008    & 0.028    \\
    \bottomrule
  \end{tabular}
  \caption{Testing Result Comparing \textbf{AdaptJobRec} with Baselines on Job Recommendation Task}
  \label{tab:pvals-job}
\end{table}

We collected user profiles, browsing histories, and click data from 10,014 users interacting with the Walmart Job Recommendation System in 2024, and subsequently compared the performance of AdaptJobRec, LLM RAG~\cite{gao2023retrieval}, the ReAct Agent~\cite{yao2023react}, MACRS Agent~\cite{fang2024multi}, and the Plan \& Execute Agent~\cite{wang2023plan} on job recommendation tasks.

Table~\ref{tab:job} shows that AdaptJobRec consistently outperforms all four baselines, across metrics Hit@10, NDCG@10, and MAP@10, achieving the highest scores of 0.3176, 0.0810, and 0.0371, respectively. Compared with the strongest baseline (Plan \& Execute), AdaptJobRec still delivers measurable improvements in all metrics. As confirmed by Welch’s $t$-test in Table~\ref{tab:pvals-job}, these gains are statistically significant, highlighting AdaptJobRec’s robust advantage in both retrieval accuracy and ranking quality for job recommendation tasks.

%(p$<$0.001 vs. RAG and ReAct; p$\leq$0.05 vs. Plan \& Execute and MACRS)
%Table~\ref{tab:job} illustrates that the agentic conversational recommendation system based on the AdaptJobRec architecture outperforms the LLM RAG method, ReAct Agent, Plan \& Execute Agent and MACRS Agent across Hit@10, NDCG@10, and MAP@10 metrics.

%\textcolor{blue}{In Table~\ref{tab:pvals-job}, we report statistical test (Welch’s t-test) results for AdaptJobRec and four baseline methods across three metrics. As shown in the results, AdaptJobRec significantly outperforms the other four methods on the job recommendation task.}

%\FloatBarrier

\subsection{Career Path Prediction}

%How to integrate external knowledge into LLMs, whether through fine-tuning methods or augmentation tools such as databases, recommendation engines, knowledge graphs, or search engines, remains a contentious issue. The Walmart People Kafka system publishes approximately 14,400 job opening updates every day. In such a rapidly changing environment, enhancing LLMs through fine-tuning methods becomes impractical. In contrast, career path development data is relatively stable; a reasonable career path does not become obsolete within just 1–2 months. Therefore, we collected 2,735,471 job transition records for Walmart associates from 2022 to 2023 and performed fine-tuning using Llama-3.1-8B \cite{touvron2023llama} and DeepSeek-R1-Distill-Qwen-7B \cite{liu2024deepseek} as bases, resulting in the Llama-Capa and DeepSeek-Capa models, for comparison with the AdaptJobRec. We used Walmart’s internal 2024 job transition dataset (comprising 2,299,008 records) as a test set to compare the performance of the AdaptJobRec against two fine-tuned LLM models, Llama-Capa and DeepSeek-Capa.

We collected 932,854 Walmart associate job transition records from 2022 to 2023 as training data for AdaptJobRec. We fine-tuned Llama-3.1-8B~\cite{touvron2023llama} and DeepSeek-R1-Distill-Qwen-7B~\cite{liu2024deepseek} on this dataset, yielding the Llama-Capa and DeepSeek-Capa models. We then used 471,495 job transition records from Walmart associates in 2024 as testing data to compare AdaptJobRec’s performance against the two fine-tuned LLM models, Llama-Capa and DeepSeek-Capa.

\begin{table}[t]
\centering
  \begin{tabular}{lccc}
    \toprule
    Methods & \% Hit Real Trans $\uparrow$ & Latency(s) $\downarrow$\\
    \midrule
    Frequency & 8.12 & 0.32 \\ 
    Llama-Capa & 8.35 & 0.67 \\
    DeepSeek-Capa & 11.24 & 0.81 \\
    {\bfseries AdaptJobRec}  & {\bfseries 12.82} & {\bfseries 0.36} \\
  \bottomrule
\end{tabular}
\caption{Comparison on Career Path Prediction Task}
\label{tab:path}
\end{table}

% Table: p-values vs fine-tuned LLMs
\begin{table}[t]
  \centering
  \begin{tabular}{@{}lcc@{}}
    \toprule
    \shortstack[c]{AdaptJobRec\\ vs. Baseline} & \shortstack[c]{$p$-value \\ (Hit Real Trans)} & \shortstack[c]{$p$-value \\ (Latency)} \\
    \midrule
    Frequency      & 0.017     & -- \\
    Llama-Capa     & 0.018  & 0.021 \\
    DeepSeek-Capa  & 0.031  & 0.025 \\
    \bottomrule
  \end{tabular}
  \caption{Testing Result Comparing \textbf{AdaptJobRec} with Baselines on Career Path Prediction Task}
  \label{tab:pvals-llm}
\end{table}

As shown in Table~\ref{tab:path} and Table~\ref{tab:pvals-llm}, AdaptJobRec attains the highest hit rate of real transitions (12.82\%) among all methods, significantly surpassing Frequency, Llama-Capa, and DeepSeek-Capa. Despite its superior accuracy, AdaptJobRec maintains a low response latency (0.36~s), significantly faster than Llama-Capa and DeepSeek-Capa and close to the minimal latency Frequency method. These results demonstrate that AdaptJobRec achieves an effective balance between accuracy and efficiency for career path prediction.

%\textcolor{blue}{Table~\ref{tab:path} and Table~\ref{tab:pvals-llm} show that the AdaptJobRec Agentic system hit a significant higher percentage of real transitions than that of the Frequency transition method~\cite{ghosh2020skill}, Llama-Capa model, and DeepSeek-Capa model, and the response latency of AdaptJobRec is significantly lower than that of Llama-Capa and DeepSeek-Capa. }

%Therefore, we selected the career path growth algorithm as the tool for our AdaptJobRec conversational career recommendation system.
%job recommendation，view，like，dislike，save，positive/negative sentimental response; user's job transition history and user saved career path recommendation; complex user queries data (Containing synthetic data)
% \url{https://ctan.org/pkg/booktabs} --- for preparing

\subsection{Pilot Group Study}

We logged 150 conversation sessions from a pilot group of 30 users. At the start of each session, the experimental UI randomly selected one of four methods: the ReAct Agent~\cite{yao2023react}, the Plan \& Execute Agent~\cite{wang2023plan}, MACRS Agent~\cite{fang2024multi}, or AdaptJobRec.  We measured the average number of conversation rounds needed to obtain the target information and the total session response time for each user.  Conversation rounds of sessions in which users hadn’t acquired the information by the end of the conversation were assigned as 20.

%In addition, the career path growth algorithm can dynamically adjust its recommendation bias based on real-time user feedback.

\begin{table}[t]
  \centering
  \begin{tabular}{lccc}
    \toprule
    Methods & Conv Round $\downarrow$ & Resp Latency (ms) $\downarrow$\\
    \midrule
    RAG & 7.10 & 1065 \\
    Plan \& Execute & 6.38 & 957\\
    ReAct & 3.86 & 579 \\
    MACRS & 3.68 & 552 \\
    {\bfseries AdaptJobRec}  & {\bfseries 3.32} & {\bfseries 498} \\
  \bottomrule
\end{tabular}
\caption{Comparison of Average Conversation Rounds and Response Latency for the Pilot Group}
\label{tab:cov_rd}
\end{table}

% Table 2: p-values on system-level metrics
\begin{table}[t]
  \centering
  \begin{tabular}{@{}lcc@{}}
    \toprule
    \shortstack[c]{AdaptJobRec\\ vs. Baseline} & \shortstack[c]{$p$-value \\ (Conv Round)} & \shortstack[c]{$p$-value \\ (Latency)} \\
    \midrule
    RAG                  & $<$0.001 & $<$0.001 \\
    ReAct                & $<$0.001 & $<$0.001 \\
    Plan \& Execute & $<$0.001 & $<$0.001 \\
    MACRS                & 0.002    & $<$0.001 \\
    \bottomrule
  \end{tabular}
  \caption{A/B Testing Result Comparing \textbf{AdaptJobRec} with Baselines on Average Conversation Rounds and Response Latency for the Pilot Group}
  \label{tab:pvals-sys}
\end{table}

%\xw{Some writings (e.g., two metrics) of this paragraph should be merged to the previous graph when describing Table 5. The two n values can be removed to reduce confusion.}

Table~\ref{tab:cov_rd} and Table~\ref{tab:pvals-sys} present the comparison and Welch’s t-test results between AdaptJobRec and four baseline methods (RAG, Plan \& Execute, ReAct, and MACRS) on two key metrics: (1) the average number of conversation rounds per session (2) the average response latency. As shown in Table~\ref{tab:cov_rd}, AdaptJobRec achieves the lowest average conversation rounds (3.32) and the fastest response latency (498 ms), representing a 54\% reduction in conversation rounds and a 53\% improvement in latency compared to the RAG baseline. Welch’s t-test results in Table~\ref{tab:pvals-sys} confirm that AdaptJobRec’s improvements over the baselines are statistically significant.

During the pilot study, users intentionally asked large percentage of complex queries to test the system, leaving simple queries at under 5\% of requests. This limited AdaptJobRec’s ability to showcase its lower response latency. After the application launch, real world usage will include many more simple queries, making its latency advantage much more apparent.

\section{Conclusion and Future Work}
In this work, we present AdaptJobRec, an LLM-powered agentic conversational job recommendation system that handles diverse career recommendation tasks. AdaptJobRec incorporates a complexity identification mechanism, a memory processing module, and a task decomposition planner with multiple personalized recommendation tools. Simple queries are routed directly to the appropriate tool for rapid responses. Complex queries are first handled by the memory processing module and then passed to the Intelligent task planner. These processes eliminates redundant planning and boosts the planner’s tool selection accuracy. Evaluation across varied scenarios shows that AdaptJobRec significantly reduces latency without sacrificing accuracy. Future work will broaden its personalized toolset and further refine each module’s performance.
%\xw{better to have 6 page content and the references start exactly at page 7. Consider to add a bit more writings of deployment architecture, baselines, or even metrics. The Planner paragraph is short and can be expanded. A paragraph of paper outline can be added at the end of Introduction if you need to fill out space. The new writings can also be reflected in abstract, introduction, and conclusion.}

%\section{Acknowledgments}
%xxxx

\bibliography{aaai2026}

\end{document}